\documentstyle[aps,pre,twocolumn,epsf]{revtex}

\begin{document}
\draft 
\tightenlines

\title{Noise-Sustained Currents in Quasigeostrophic Turbulence over 
Topography}

\author{Alberto Alvarez$^*$, Emilio Hern\'andez-Garc\'\i a$^{*\#}$, 
 and Joaqu\'\i n Tintor\'e$^{*\#}$}
\address{$^*$Departament de F\'\i sica, Universitat  
de les Illes Balears, E-07071 Palma de Mallorca, SPAIN \\
$^\#$ Instituto Mediterr\'aneo de Estudios Avanzados, IMEDEA (CSIC-UIB)\\  
 E-07071 Palma de Mallorca, Spain\\
URL: http://www.imedea.uib.es/}
\date{\today}
\maketitle

\begin{abstract}
We study the development of mean structures in a nonlinear model of large
scale ocean dynamics with bottom topography and dissipation, and forced with a
noise term. We show that the presence of noise in this nonlinear model leads
to persistent average currents directed along isobaths. At variance with
previous works we use a scale unselective  dissipation, so that the phenomenon
can not be explained in terms of minimum enstrophy states. The effect requires
the presence of both the nonlinear and the random terms, and can be though of
as an ordering of the stochastic energy input by the combined effect of
nonlinearity and topography. The statistically steady state is well
described by a generalized canonical equilibrium with mean energy and
enstrophy determined by a balance between random forcing and dissipation. This
result allows predicting the strengh of the noise-sustained currents. 
Finally we discuss the relevance that these noise-induced currents could have
on real ocean circulation. 
\end{abstract}


\section{Introduction}

Under certain conditions, nonlinear interactions are able to organize  random
inputs of energy into directed motions \cite{maddox94a,doering94}. These
ordered fluxes or structures supported by noise have important effects on 
specific  problems in biology \cite{maddox94b,douglass93}, engineering
\cite{bezrukov95} or physics \cite{rosselet94,convective}.  In fluid dynamics
noise appears in a variety of contexts. Thermal fluctuations give rise to
Langevin source terms in the equations of motion \cite{landau,thermaln}, while
stirring forces can be modeled as stochastic terms forcing the fluid motion
\cite{stirring}. Noise terms have also been  used in modeling short-scale
instabilities such as the baroclinic instability  in geophysical problems
\cite{williams78,edwards64,leith71}.  A different way in which  noise can enter
the description of fluid systems is by  considering the interactions taking
place at scales below the resolution  of a computer model.  Addition of noise
is a (crude) way of including  some influence from unresolved scales
\cite{williams78,mason94}. 

The influence of small noise on large scale fluid dynamics, far
enough from hydrodynamic instabilities and phase transitions,  
is expected to be small. However we address in this paper a situation 
in which it has important consequences: the coupling of 
fluid to bottom topography in ocean dynamics, which is one of the most
important but poorly understood factors controlling ocean circulation. We will 
show that, when interacting with noise, this coupling leads to the 
development of steady currents, even in the absence of steady
forcing. Although our results are derived for a particular quasigeostrophic 
fluid model, we expect our general conclusions to apply to other 
problems in which fluid motion occurs on irregular geometries in the 
presence of noise. 

As a first step the work in this paper will be restricted to 
barotropic, randomly forced, two-dimensional quasigeostrophic turbulence.
Several
reasons have been considered for this selection. First of all, this framework
provides a physical model complex enough to contain some relevant aspects of
ocean circulation, but easy enough for allowing physical understanding 
without 
the need of too complex numerical experiments (always difficult to interpret).
Second, baroclinic nature of ocean circulation can be usefully represented for
large scales by randomly forced barotropic models \cite{williams78}. In these
barotropic models, the random forcing formulation is a simple Markovian process
that provides a reasonable representation of the baroclinic nonlinear
interactions \cite{edwards64,leith71}. 

A brief summary of known results relevant to our study follows for
completeness. In the last decades much effort has been devoted towards 
understanding the
effects of bottom topography in barotropic models. Most relevant results 
on this
subject can be summarized in the several key works starting by that developed by
Salmon et al. \cite{salmon76} where the statistical properties of topographic
turbulence neglecting dissipation and forcing were studied. Among their results
the demonstration of the tendency to reach a maximum entropy (Gibbs) state 
characterized by
the existence of stationary mean currents following isobaths should be stressed.
The role of dissipation was analyzed in \cite{bretherton76} using
a different approach. In their study viscous damping was hypothesized, due to
its
selective damping of high wavenumber modes, to drive the flow towards a state of
minimum potential enstrophy for a given energy, hence dissipating enstrophy more
rapidly than energy which remains almost constant for a long time. 
This long-lived minimum potential
enstrophy state also consists of flow along topographic contours. More
recently, stability properties of minimum enstrophy and maximum entropy states
and their relationship have been analyzed by Carnevale and
Frederiksen \cite{carnevale87}. 
A quantitative theory of time-dependent, forced and dissipative flow over
topography was analyzed by Herring \cite{herring77} who followed the direct 
interaction approximation (DIA) \cite{kraichnan59} and a modified form of the
test-field-model turbulent formalisms \cite{kraichnan71}. He found that in the
case of random stirring and viscous topographic turbulence the stationary
components of the energy spectrum behaved close to the above mentioned inviscid
results of Salmon et al. \cite{salmon76}. A statistical theory simpler than DIA
was used in \cite{holloway78} to investigate the time evolution of the energy
and vorticity-topography correlation spectrum. His results also show the
emergence
of steady circulation for a certain range of the parameters of the theory. 
As a first step towards understanding the more complicated 
baroclinic case, Treguier 
\cite{treguier89} made an extensive exploration of the parameter space for
barotropic turbulence. Her numerical results show a non-monotonic dependence of
the percentage of steady currents on the forcing, viscous and topography
parameters. In some simulations where random forcing and dissipation were
considered, these steady currents resemble the simple inviscid solution 
found in \cite{salmon76}.

 All these previous studies address the existence of stationary currents
following isobaths in a variety of conditions involving random noise. However 
from these studies it is not possible to infer a clear relation between noise
and the physical nature of these currents. The reason is that
scale selective (i.e., viscosity) dampings were considered so that 
the generation of directed currents can be explained in terms of minimum 
enstrophy states \cite{bretherton76}. This physical mechanism
involving minimum enstrophy solutions does not require the existence
of noise. 

 In this paper, we try to highlight 
the role of noise in the appearance of steady 
currents in such open systems. The main difference with previous studies is 
that we have considered topographic turbulence damped by Rayleigh friction
(bottom friction) instead of viscosity damping. This introduces important
physical differences because Rayleigh friction provides  
a scale-unselective way of dissipating energy from the system and thus 
does not produce transient minimum enstrophy states. 
In this way, avoiding
one of the possible mechanisms for explaining the mean flows that were  
present in the simulations of \cite{herring77},\cite{holloway78} and 
\cite{treguier89}, we show that the combined effect of noise and nonlinear 
terms is enough to generate mean flows when bottom topography is present. 
In order to get a more physical understanding of this phenomenon,
we show that this current formation at large scales is 
described, to a good degree of approximation, by a 
generalized canonical distribution in which the mean enstrophy and the 
mean energy are fixed by the balance between random forcing and dissipation. 
Then we suggest that the generation of currents by noise in this system 
can be understood as 
arising from the tendency of the large scale fluid motions in our 
model to approach 
equilibrium-like states, that is states with the maximum possible entropy 
consistent with given constraints. 
However, the non-equilibrium character of our system is reflected in the small 
scales, which depart from this equilibrium-like state.  

Two comments are in
order: First, since our motivation is the understanding of large scale
ocean circulation, we are interested in the final asymptotic behavior of 
the fluid
system under forcing and dissipation, and not in the approach to it. Therefore
the numerous studies available for freely decaying turbulence can not be applied
to this problem. Second, the presence of topography introduces spatial scales
into the problem so that mechanisms related to scale-invariant wavenumber
cascades \cite{carnevale92} can not be as relevant here as in a flat
bottom situation. 

The paper is organized as follows: 
in Section \ref{model} the quasigeostrophic model is introduced and
numerically solved to show the appearance of directed currents along the
topography. Section \ref{analytic} contains analytical considerations showing 
the relevance of the nonlinear terms, discussion of the Fokker-Planck
equation for the model and calculation of some average values. The
generalized canonical distribution is also presented and some consequences
are derived from it. Section \ref{comparison} checks the validity of the
canonical description. Discussion of the results and their implications for
large-scale ocean modelling are in Section \ref{discussion}.

\section{The model and the Noise-Sustained Currents}
\label{model}

The quasi-geostrophic evolution equation for the streamfunction $\psi(x,y,t)$ 
describing flow over topography is given by \cite{pedlosky87}:
\begin{equation}
\label{QG}
{\partial \nabla^2 \psi \over \partial t} +
{\cal J}  \left( \psi , \nabla^2 \psi +h' \right) =
-\epsilon \nabla^2 \psi + F \ , 
\end{equation}
where 
\begin{equation}
{\cal J}  \left( A , B \right) \equiv
{\bf \hat k} \cdot \left(\nabla A \times \nabla B  \right)
\end{equation}
$\epsilon^{-1}$ is the bottom friction decay time, $h'=f \Delta H/H_0$
with $f$ the Coriolis parameter (that will take the value $10^{-4} s^{-1}$ as 
appropriate for mean latitudes), 
$H_0$ the mean depth, and $\Delta H(x,y)$ the
local deviation from this mean depth. 
${\bf \hat k}$ is the unit vector normal to the horizontal
plane $(x,y)$, $\times$ denotes the cross vector product and $F(x,y,t)$ 
is a Gaussian white-noise process  with zero
mean and
correlations $\left< F(x,y,t)F(x',y',t') \right> =
D \delta (x-x')\delta (y-y')\delta (t-t')$.
It represents relative-vorticity random forcing. Eq.(\ref{QG}) is to 
be solved in a square domain with boundary conditions such that $\psi$ 
vanishes at the boundaries. 
The streamfunction $\psi$ gives the horizontal components of the fluid 
velocity $\left( u(x,y),v(x,y) \right)$ as 
\begin{equation}
u = - {\partial \psi \over \partial y}\ , \ \ \ 
v =  {\partial \psi \over \partial x}
\end{equation}

In order to solve (\ref{QG}) we have used the quasi-geostrophic 
numerical scheme developed in  
\cite{cummins92} on a grid of $128 \times 128$ points. The resolution 
of our scheme (corresponding to the distance  
between grid points) is of 10 km, so that the total system 
size is $L=1280$ km. The algorithm,
based on finite differences and the leap-frog algorithm, keeps the value of
energy and enstrophy constant when it is run in the inviscid and unforced case.
The consistent way of introducing the stochastic term into the leap-frog  
scheme can be found following the lines of \cite{sancho82,greiner88}:  
Basically, a field of Gaussian random numbers is generated at each time step,
and the random field added to the deterministically evolved vorticity field at
each time step is the sum of the field generated at this time step plus the
field generated at the previous time step. The random number generator employed
in the simulation is detailed in \cite{toral93}.  The amplitude of the forcing,
$D=2\times 10^{-9}\  m^2/s^2$, and the $e$-folding  decay constant, $\epsilon
=3 \times 10^{-7} s^{-1}$ of the damping mechanism  have been chosen in order
to obtain final velocities of several centimeters per second. The topographic
field (Fig 1) is randomly generated from a isotropic spectrum containing, with
equal amplitude and random phases, all the Fourier modes corresponding to 
scales
between 80 km and 300 km. The model was run for $5 \times 10^5$ time  steps
(corresponding to 126 years) after a statistically stationary state was
reached. The average streamfunction was then computed  from the configurations
obtained every time step.

The results obtained from this numerical experiment show the generation of a
stationary mean flow correlated with bottom topography and with velocity values
of order of 3 cm/s (Fig. 2). This numerical result indicates the existence of
a physical mechanism, different from the minimum enstrophy tendency 
addressed in \cite{bretherton76}, that is able to organize
random inputs of energy in a mean state different from rest characterized by
currents with a clear dependence with bottom topography. The mean 
currents will advect any neutral tracer with them, so that the kind 
of motion so generated is very similar to the phenomena described in 
\cite{maddox94a,doering94,maddox94b,douglass93,bezrukov95,rosselet94}.  
In addition we have checked by Fourier analysis that the temporal 
series of $\psi(x,y,t)$ at fixed space point $(x,y)$ is  
roughly periodic in time (but quite noisy), so that there are also directed 
currents in the phase space of $\psi$. In the next section we try to 
clarify the possible origin of the currents. 

\section{Analytic Approaches}
\label{analytic}

\subsection{The linear model}

As a first step to investigate the emergence of these noise induced currents,
we will explore the solutions of the linearized dynamics. Neglecting the
nonlinear term, Eq. (\ref{QG}) transforms into:
\begin{equation}
\label{linear}
{\partial \nabla^2 \psi \over \partial t} +
{\bf \hat k} \cdot \left(\nabla \psi \times \nabla h'  \right) =
-\epsilon \nabla^2 \psi + F \ , 
\end{equation}	
After ensemble averaging Eq. (\ref{linear}) over noise realizations 
we obtain the equation that determines the
stationary state of the linear system:
\begin{equation}
\label{linsteady}
{\bf \hat k} \cdot \left(\nabla \left< \psi \right> \times \nabla h'  \right) 
+ \epsilon \nabla^2 \left< \psi \right> = 0
\end{equation}

No stationary solution of Eq. (\ref{linsteady}) can be a
function of topography. This is because a functional dependence of the mean
streamfunction on the bottom topography implies that the first term of 
(\ref{linsteady}) would
be identically zero and then $<\psi>$ should be a solution of 
Laplace's equation. With the boundary conditions used, this implies 
that the mean streamfunction vanishes in every
point of the system. Thus in the linear model (\ref{linear}) the
forcing generates small excursions of $<\psi>$ around the rest state, 
in a way unrelated to topography.  
In conclusion, the nature of noise sustained currents is thus determined 
not only by the random noise but also by the nonlinear 
interactions.

\subsection{The Fokker-Planck equation of the nonlinear model}

As it was shown from the study of the linearized dynamics,
the nonlinear term is needed in order to get mean stationary solutions different
from rest. This fact implies to consider the full dynamics given by (\ref{QG}). 
Due to
the stochastic nature of the system we will explore the statistics of its
solutions in phase space. To do that, the Fokker-Planck equation (FPE) of the
probability distribution of the field solution of Eq. (\ref{QG}) will be 
derived. From the
standpoint of treating the statistics of an ensemble of solutions of 
(\ref{QG}), it is
convenient to represent the streamfunction $\psi(x,y,t)$ as a linear 
combination of
orthogonal eigenfunctions of the Laplacian satisfying the adequate 
boundary conditions:
\begin{equation}
\label{decom}
\psi(x,y,t)=\sum_i A_i(t) \phi_i(x,y), 
\ \ \nabla^2 \phi_i(x,y)=-\alpha^2_i \phi_i(x,y)
\end{equation} 

\begin{equation}
\int_S dx dy \phi_i(x,y)\phi_j(x,y)= S \delta_{ij}
\end{equation} 
The integral is over the whole two-dimensional domain, of area $S=L^2$.

The evolution equation for the amplitude factors $A_i(t)$ reads:
\begin{equation}
\label{adot}
\alpha_k^2 \dot A_k - 
\sum_{ij}\beta_{ijk}\left(\alpha_j^2 A_j A_i -A_i h'_j\right)=-\epsilon
\alpha_k^2 A_k + \alpha_k f_k
\end{equation}
where:
\begin{equation}
\alpha_k f_k= - {1 \over S}\int_S dx dy \ F(x,y,t) \phi_k(x,y) \ , 
\end{equation}

\begin{equation}
\label{ff}
\left< f_k(t) f_{k'}(t')\right> =
 {D \over S \alpha_k \alpha_{k'}}\delta_{kk'}\delta (t-t')\ , 
\end{equation}
and 
\begin{equation}	
\beta_{ijk}= - {1 \over S}\int_S dx dy \ {\bf \hat k} \cdot 
\left(\nabla \phi_i(x,y) \times \nabla \phi_j(x,y)  \right) \phi_k(x,y) \ . 
\end{equation}

Some properties of the nonlinear interaction coefficients $\beta_{ijk}$ 
are that they  
vanish if any two indexes are equal, they are invariant under cyclic 
permutation
of indices, and they reverse sign under non-cyclic permutation of indices
\cite{thompson72}.

In order to use the symmetries and asymmetries of Eq. (\ref{adot}) more 
clearly, we introduce the following change of variables: 
\begin{equation}
\label{X}
X_k(t) \equiv \alpha_k A_k(t)\ , \ h'_j \equiv \alpha_j h_j\ . 
\end{equation}
The equation is now:
\begin{equation}
\label{Xdot}
\dot X_k(t) = N_k - \epsilon X_k(t) + f_k(t) 
\end{equation}
where:
\begin{equation}
\label{N}
N_k = \sum_{ij} {\beta_{ijk} \alpha_j^2 \over \alpha_j \alpha_i \alpha_k}
\left( X_i X_j - X_i h_j \right) 
\end{equation}
 Now, we consider the amplitudes $\{X_i\}$ of the orthogonal 
modes as the coordinates
of a point (in an infinite-dimensional phase space) whose position corresponds
to the state of a particular realization at a single instant in time. Since 
$f_k(t)$ is white noise (Eq. (\ref{ff})), the continuity equation 
for the phase space probability density $\rho=\rho(\{X_i\},t)$ takes the
Fokker-Planck form \cite{gardiner89}:
\begin{equation}
\label{FP}
{\partial \rho \over \partial t} + 
\sum_k {\partial \over \partial X_k} \rho \left( N_k - \epsilon X_k \right) = 
\sum_k \mu_k {\partial^2  \rho \over \partial X_k^2}
\end{equation}
where $\mu_k = {D / (2 S \alpha_k^{2})}$ and represents a diffusion
coefficient for the $k-$mode. We can see from (\ref{FP}) that random forcing 
acts as a
diffusion mechanism for the probability density in phase space. This mixing
character is opposite to the term represented by friction that tends to drive
the system towards the origin of the phase space, the rest state.

The FPE (\ref{FP}) can be easily
integrated in the stationary state if the potential condition:
\begin{equation}
\label{potential}
{\partial Z_k \over \partial X_l} = {\partial Z_l \over \partial X_k}
\end{equation}
with  $Z_k = \mu_k^{-1} \left( N_k -\epsilon X_k \right)$ holds 
\cite{gardiner89}. Otherwise there is no simple way of finding the
stationary probability distribution. The potential condition is not satisfied 
in the case described by (\ref{FP}) so that some assumptions must be 
performed to get an
approximate probability distribution for the stationary state. 

However the FPE can be still useful for determining some average quantities
needed below. The average of any phase space 
function $F(\{X_k\})$ satisfy the adjoint of Eq.(\ref{FP}), or {\sl backwards} 
equation \cite{gardiner89}. This can be shown by multiplying (\ref{FP}) by  
$F(\{X_k\})$, integrating over phase space, and using integration by parts. 
For stationary averages (${\partial \rho \over \partial t} = 0$), the result is 
\begin{equation}
\label{backwards}
\int d\Gamma \left( {\partial F \over \partial X_l} 
\left(N_k - \epsilon X_k \right) + \mu_k {\partial^2 F \over \partial X_k^2}
\right) \rho = 0
\end{equation}
$\int d\Gamma $ denotes integration over phase space $\int \prod_k dX_k $. 
Choosing $F = {1 \over 2} \alpha_k^2 \left({1 \over 2} X_k - h_k \right)^2$,
substituting into (\ref{backwards}) and summing over $k$, the antisymmetric 
terms containing $\beta_{ijk}$ disappear and we find:   
\begin{equation}
\label{aveQ}
\epsilon \sum_k \alpha_k^2 
\left< \left( {1 \over 2} X_k -h_k \right) X_k \right> = 
\sum_k {1 \over 2} \mu_k \alpha_k^2
\end{equation}
where $<...>$  indicates average over the stationary distribution $\rho$. 
The term between brackets is related to the mean potential enstrophy, 
so that Eq.
(\ref{aveQ}) gives an expression for this quantity in terms of the 
damping coefficient $\epsilon$ and the noise intensity $\mu_k$. 
Similarly, by choosing $F={1 \over 4}X_k^2$ and repeating the procedure 
above we find:
\begin{equation}
\label{aveE}                      
\epsilon \sum_k \alpha_k^2 
\left< {1 \over 2} X_k^2 \right> = 
\sum_k {1 \over 2} \mu_k 
\end{equation}
This relates the mean energy $\left< {1 \over 2} X_k^2 \right>$ to  
$\epsilon$ and $\mu_k$.

\subsection{A generalized canonical equilibrium} 

Properties of the stationary state of the FPE (\ref{FP}) can be used to obtain 
an
approximate expression for the probability distribution. 
As it was shown in \cite{salmon76}, invariance of energy and potential
enstrophy has important implications on the inviscid and unforced dynamics. In
particular statistical properties can be correctly computed by assuming that at
long times the system explores completely, or at least samples with enough
significance, the hypersurface of constant energy and potential enstrophy  
(for a discussion see \cite{jup}). 
When damping is introduced, the system will be brought to the vicinity of the
rest state in phase space. But random forcing will keep some motion present.
Since the effect of the nonlinear terms in the unforced and undamped dynamics
was to drive the system along the intersection of the constant-energy and 
constant-potential-enstrophy
hypersurfaces, it is then reasonable to expect that in the presence of forcing
and damping evolution will proceed on average close to the 
hypersurfaces of mean
constant energy and mean potential enstrophy determined by the damping and
forcing coefficients through equations (\ref{aveQ}) and (\ref{aveE}). If this 
hypothesis holds,
the most likely probability distribution compatible with such constraints, that
is the one obtained from the maximization of the entropy functional 
$S[\rho] = \int \prod_k dX_k \rho(\{X_k\}) \log \rho(\{X_k\})$ 
subjected to the constraints (\ref{aveQ}) and (\ref{aveE}), should be an 
approximation to the
real stationary probability distribution \cite{maxent}. 

The validity of the heuristic considerations above, as the validity of general
statistical mechanics arguments in any interesting physical situation, would be
hard to establish rigorously. In our case, justification of our assumptions will
come from comparisons with numerical simulations. Ergodicity, or at least
efficient sampling of the constant energy and potential enstrophy 
hypersurfaces would be certainly more justified here than in the unforced case 
because of the
additional mixing tendency provided by random forcing. 

Entropy maximization subjected to constraints (\ref{aveQ}) and (\ref{aveE}) 
leads formally to the same expression as in the
inviscid and unforced case, that is, the predicted probability distribution is:
\begin{equation}
\label{prodrho}
\rho(\{X_k\}) = \prod_k \rho_k(X_k) , 
\end{equation}
with
\begin{equation}
\label{canonical}
\rho_k(X_k) \ = e^{\left( -{1 \over 2} \left( a +b\alpha_k^2 \right) X_k^2 
+ \alpha_k^2 h_k b X_k \right)}
\end{equation}
$a$ and $b$ are the Lagrange multipliers enforcing the constraints. Although
this expression is formally identical to the inviscid one, there is an important
difference: In the inviscid case the Lagrange multipliers are
determined from the initial values of energy and enstrophy, whereas here the
dependence on the initial state is asymptotically lost and 
(\ref{canonical}) refers to the
final state in which the Lagrange multipliers are determined by the mean 
values of energy and enstrophy. Their exact values are given in (\ref{aveQ}) and
(\ref{aveE}) in terms of the forcing and friction
parameters present in the original equation (\ref{QG}). 
The values of the first and second moments of $X_k$ obtained from 
(\ref{canonical}) are given by:
\begin{equation}
\label{aveX}
\left< X_k \right> = {\alpha_k^2 h_k b \over a + b \alpha_k^2 }
\end{equation}
\begin{equation}
\label{aveX2}
\left< X_k^2 \right> = {1 \over a + b \alpha_k^2 }+
{\alpha_k^4 h_k^2 b^2 \over a + b \alpha_k^2 }
\end{equation}

These equations in conjunction with (\ref{aveQ}) and (\ref{aveE}) 
allow the determination of
the Lagrange multipliers $a$ and $b$ from the parameters in the FPE or in
(\ref{QG}). Alternatively, they relate the Lagrange multipliers with 
quantities measurable from simulation or experiments. 

As in the inviscid case, equations (\ref{aveX}) and (\ref{aveX2}) imply the 
existence of a mean circulation controlled by
the topography \cite{salmon76}. To see that we focus on the value of the
first moment which is specially interesting since it describes the generation 
of mean flows following isobaths. This becomes evident rewriting (\ref{aveX})   
in terms of the Fourier amplitudes of the functional expansion (\ref{decom}):
\begin{equation}
\label{fourier}
\left< A_k \right> = {h_k' b \over a + b \alpha_k^2 }
\end{equation}
that in the physical space imply the relation:
\begin{equation}
\label{physical}
\nabla^2 \left< \psi \right> + h' = c \left< \psi \right>
\end{equation}
where $c=a / b$. This
linear relationship between the mean potential vorticity and the mean
streamfunction implies, considering that for large scale motions the relative
vorticity is smaller than the ambient one 
($\nabla^2 \left< \psi \right> \ll h' $), that $\left< \psi \right>$ 
is proportional to $h'$, that is $\left< \psi \right>$ describes mean flows
following the isobaths. 
We have checked that the distribution 
(\ref{prodrho})-(\ref{canonical}) is not an exact solution of the 
FPE (\ref{FP}). In the following Section however we show that it 
gives a good description of the large scale average properties. 

\section{Comparison with Numerical Simulations}
\label{comparison}

The correlation between topography and mean streamfunction, as 
predicted by (\ref{physical}) is obvious by comparing Figs. 1 and 2. 
The agreement can be
made more quantitative by plotting $\nabla^2 \left< \psi \right> + h'$ 
versus $\left< \psi \right>$, both obtained from the numerical 
simulation. This is done in Fig 3. Eq. (\ref{physical}) 
predicts a linear relationship. Fig. 3 shows a clear linear trend but 
there is also an evident scatter in the data that will be commented 
below. The linear correlation coefficient is 0.940. Further analysis of the 
validity of theoretical predictions
can be done by fitting the data in Fig. 3 to Eq. (\ref{physical}) to get
a numerical value for the constant $c$. This procedure gives a value of $c= 
(5.30  \pm 0.02) \times 10^{-9}\ m^{-2}$. The predicted value 
of $c$ obtained from (\ref{aveX}) and (\ref{aveX2}) by considering the 
magnitudes of the
mean energy and potential enstrophy obtained in the stationary state is 
$5.5 \times 10^{-9} m^{-2}$ which represents a deviation of $4\%$ with respect 
to the value obtained
by numerical simulation. In order to test if such deviations occur at all
scales, we have computed the power spectrum of the fields 
$(\nabla^2 \left< \psi \right> + h')/c$ 
and $\left< \psi \right>$, both obtained from the numerical model,
and compared them with the energy spectrum of the field 
$\left< \psi \right>$ computed from direct 
resolution of (\ref{physical}) which will be taken as reference spectrum. 
Theory predicts that these
distributions should be the same. Fig. 4 shows that there is very good agreement
between numerical results and the theoretical spectrum obtained from the
canonical probability distribution at low wavenumbers (large scales) where 
topography components are relevant. The most important 
deviations occur at small scales. There are also much smaller discrepancies 
at very low wavenumbers, that is at basin scales, similar to the ones observed 
when analyzing the equilibrium solutions of purely inviscid two-dimensional
turbulence \cite{dev}. The reason for them could be the long dynamical time 
scale of large-scale eddies that implies very long equilibration times 
\cite{dev}. The deviations involving
small scales cannot be explained by relaxation-time arguments. These scales 
are clearly out of equilibrium. This absence of equilibrium at 
least at small scales 
could be expected from arguments based on turbulent cascades: In
two-dimensional turbulence an enstrophy flux towards small scales is expected
\cite{turbulence2d}. Since Rayleigh friction is not as effective in dissipating
vorticity at small
scales as viscosity would be, an accumulation of enstrophy is expected at the
largest wavenumbers, and this is in fact seen in Fig. 4. In contrast, equations   
(\ref{prodrho})-(\ref{canonical}) predict that states with vorticity at 
wavenumbers above a cut-off are forbidden. This is easy to see from the
canonical partition function of the system, that can be written as:
\begin{equation}
\label{Z}
Z(b,a)=\int {\cal D} \psi\ 
e^{- {a \over 2}( \int dx dy \left( \nabla \psi \right)^2
   + {l^2}\int dx dy \left( \nabla^2 \psi +h' \right)^2)   }
\end{equation}
where $l^{2}= b/a=c^{-1}$ has units of length square and 
$\int {\cal D} \psi$
denotes functional integration. If $\nabla^2 \psi +h'$ 
varies on length scales smaller than $l$  the second
term of the exponential dominates and all the microstates of the system with
these characteristics are dropped out from the ensemble averages. As a result 
properties of scales near or smaller than the cut-off cannot be well described
by the canonical probability distribution.

The fact that small scales are responsible for the broadening of the linear
tendency in Fig. 3 can be verified from the fact that the linear correlation
coefficient between  mean potential vorticity 
and mean streamfunction fields after 
scales smaller than $100$ km (i.e. several times the cut-off scale) have been
filtered out is 0.975, higher than the original one. Finally, if all scales 
except those with relevant topographic
components ($100-400$ km) are removed, correlation coefficient increases until 
a value of 0.982, that is, the agreement with theory predictions is even better. 
In conclusion the canonical distribution given by 
(\ref{prodrho})-(\ref{canonical}) 
is a good description of the numerical results for length scales 
larger than $100$ km. 

\section{Discussion}
\label{discussion}

	In this
paper, we have focussed our attention on the effects that noise produces
on a simplified stochastic
quasi-geostrophic model where only friction and random terms have been
considered.  The reason for our interest comes from indications of
generation of currents along isobaths in randomly forced quasigeostrophic 
systems presented in previous works 
\cite{herring77,holloway78,treguier89}. Since the main interest
of those works was not the study of the effect of random forcing on flow with 
underlying topography, the formulations used there included several effects 
absent in our model. In particular numerical experiments were carried out 
in the presence of correlated noise and
scale-selective dampings. This more complex framework makes it difficult to 
identify unambiguously the basic physical mechanisms producing the currents. 
In particular the introduction of viscosity or bi-laplacian damping 
generates decaying minimum enstrophy states described by \cite{bretherton76}
that obscure the effects of noise highlighted in our work.  Our use of 
bottom friction as a damping mechanism tries to isolate 
the effect that noise could produce
on the dynamics by avoiding minimum enstrophy states. 

	As a result of our study, we have found that the random term 
supplies energy to the system and
contributes (together with the nonlinear term) to the mixing of the
probability density in phase space. This compensates the reducing volume
tendency of the dissipation term. When this compensation occurs the system 
dynamics resembles the behaviour developed by the isolated system, that is
without forcing and damping. The physical effect of this phase space dynamics
is the generation of a stationary state 
characterized by mean fluid motions following isobaths. We can conclude then 
that these mean currents are noise-sustained and come from the interaction of 
random
noise with the nonlinear term. The existence of bottom topography is relevant 
in order to get a  mean component in the spectrum, that is, to develop 
mean currents in the basin. This result seems to indicate
that noise could play a relevant role in maintaining the mean component of planetary 
motions when bottom topography is present. A remaining open question is however,
how much of the
currents observed in \cite{herring77,holloway78,treguier89} is produced by 
random terms and how much is related to the minimum enstrophy tendency that is 
certainly present in their models. Our isolation of the effects of noise is a
first step to evaluate the relative importance of both effects. 

	One of the state of the art questions on the study of large scale ocean
circulation and modeling is related to the effects that unresolved scales might
have on the larger scale dynamics. The range of scales which numerical models can
handle explicitly is limited by the actual computer power, the large size of the
global domain and the time characteristics of climatic studies (long-time
simulations). These limitations prevent us from simulating the small and large  
scale features simultaneously. However, it is also well known that some of these
unresolved processes produce significant influences on the large scale
circulation. The influence of small scales must then be introduced through some
parametrization.

A natural way of considering the unresolved
dynamics would be to include the small and fast oceanic processes as some kind
of random noise in the deterministic models of large scale phenomena. In 
general, dealing 
with ocean modeling the action of small scales is only
considered as an enhancement of the viscosity coefficient (the so called
eddy-viscosity). Random noise is usually neglected from oceanic calculations
because it is assumed that its presence produces a negligible effect on 
large scale ocean dynamics.

	Although the full ocean dynamics is characterized by the existence of
three-dimensional motions, for sufficiently large spatial scales (scales 
larger than several hundreds of kilometers) and long time behaviour (more than 
one week) the 
ocean can be well described by quasigeostrophic dynamics including bottom
topography. The results shown here could be a first basis from which
to conjecture that
the origin of some observed large scale ocean structures are due to 
phenomena which
are several orders of magnitude smaller than the structures themselves. 
The experimental 
determination of this conjecture can be very
difficult to carry out due to the broad range of spatial and temporal scales  
in
ocean. However, new approaches for future studies
where large scale stochastic primitive equation models are used, instead of 
purely deterministic 
ones, could be a reasonable way to go deeper into the problem.  Effectively,
inclusion of noise in a coarse-resolution primitive equation model as a 
representation of the small scale processes that are not well resolved 
would provide a mechanism for the model to generate  
noise-sustained
large scale currents along isobaths in a natural way, 
that obviously would be missing in a 
deterministic 
version of such a model. Ad-hoc inclusion of currents along isobaths has been 
already shown to improve results of coarse-resolution numerical models 
\cite{aad,eby}. Presently, general ocean circulation models are 
deterministic. 
However, based on the results established on this paper it seems reasonable to 
hypothesize that stochastic models will provide more realistic answers
about ocean dynamics and more concretely about the possible existence and 
relevance of these noise-sustained currents in the real ocean.

\acknowledgments
Financial support from CICYT (AMB95-0901-C02-01-CP and MAR95-1861), DGICYT 
(PB94-1167 and PB94-1172), and from the MAST program 
MATTER MAS3-CT96-0051 (EC) is greatly acknowledged.

\begin{figure}
\caption{Depth contours of a randomly generated bottom topography. Maximum 
depth is 381.8m and minimum depth -381.8m over an average depth of 5000m. 
Levels are plotted every 63.6 m. Continuous contours are for positive 
deviations 
with respect to the mean, whereas dashed contours are for negative ones. }
\end{figure}

\begin{figure}
\caption{Mean streamfunction computed by time averaging when a 		   
stationary state has been achieved. Maximum and minimum values are
1120.2 and -1120.2 $m^2/s$. Levels are plotted every 186.7 $m^2/s$. Continuous 
contours denote positive values of the streamfunction, whereas dashed contours 
denote negative ones. The resemblance to Figure 1 is evident. }
\end{figure}

\begin{figure}
\caption{Scatter plot of mean potential vorticity versus mean 
streamfunction obtained from numerical simulation. A gross
linear tendency appears that resembles the
linear tendency observed in inviscid calculations. Each symbol is obtained from
a different position in the simulation domain. }
\end{figure}

\begin{figure}
\caption{Comparison, as a function of the radial wavenumber index
$L\alpha_k/2\pi$,  of the power spectra of  
$(\nabla^2 \left< \psi \right> + h')/c$ 
(dash-dot line), 
$\left< \psi \right>$ (dashed line), both obtained from  
numerical simulation, 
and the power spectra of the solution $\left< \psi \right>$ of 
Eq. (\ref{physical}) (solid line). }
\end{figure}

\onecolumn

\newpage
\ \\
\epsfxsize=14cm
\epsfysize=14cm
\epsffile{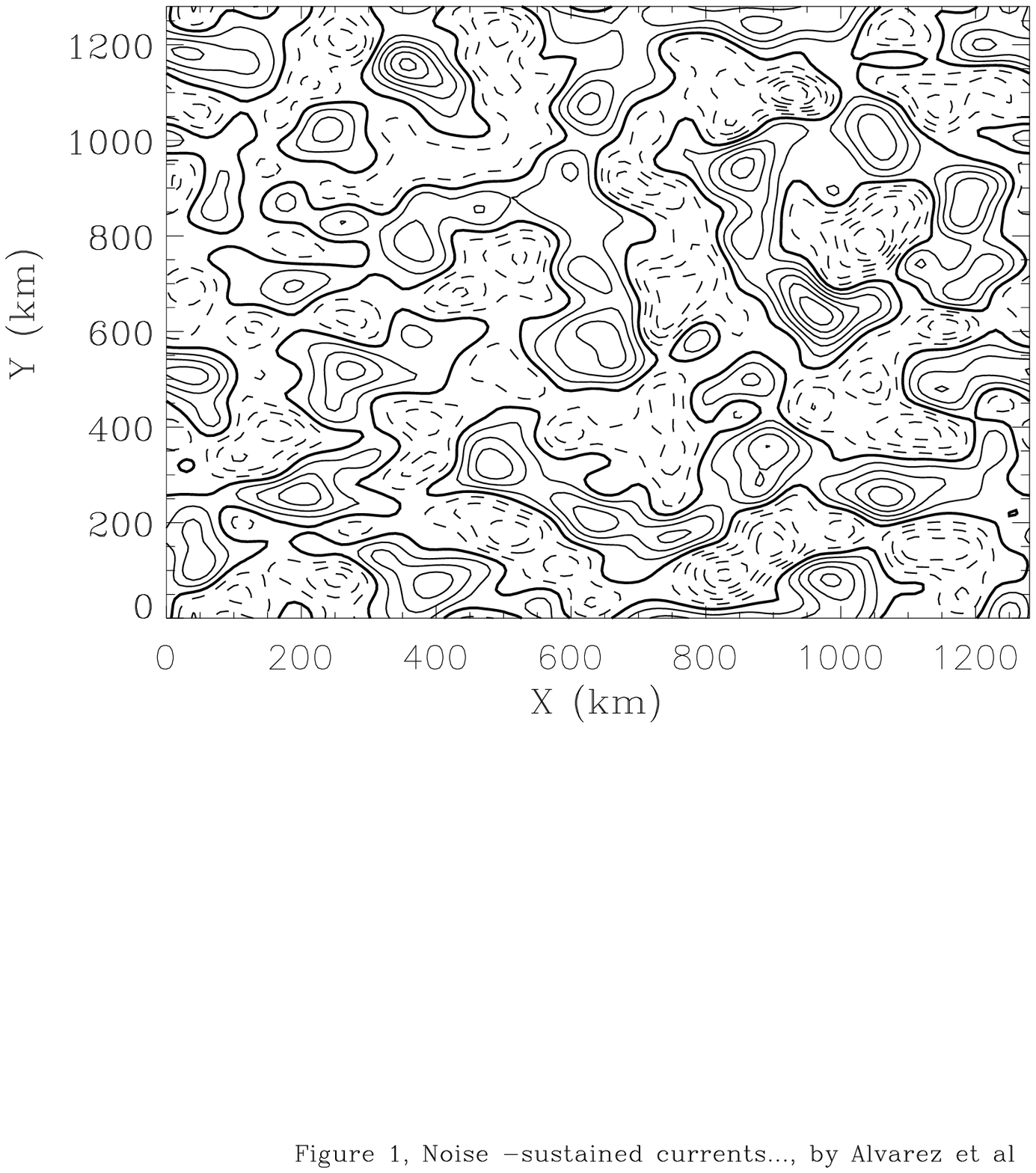}
\newpage
\epsfxsize=14cm
\epsfysize=14cm
\epsffile{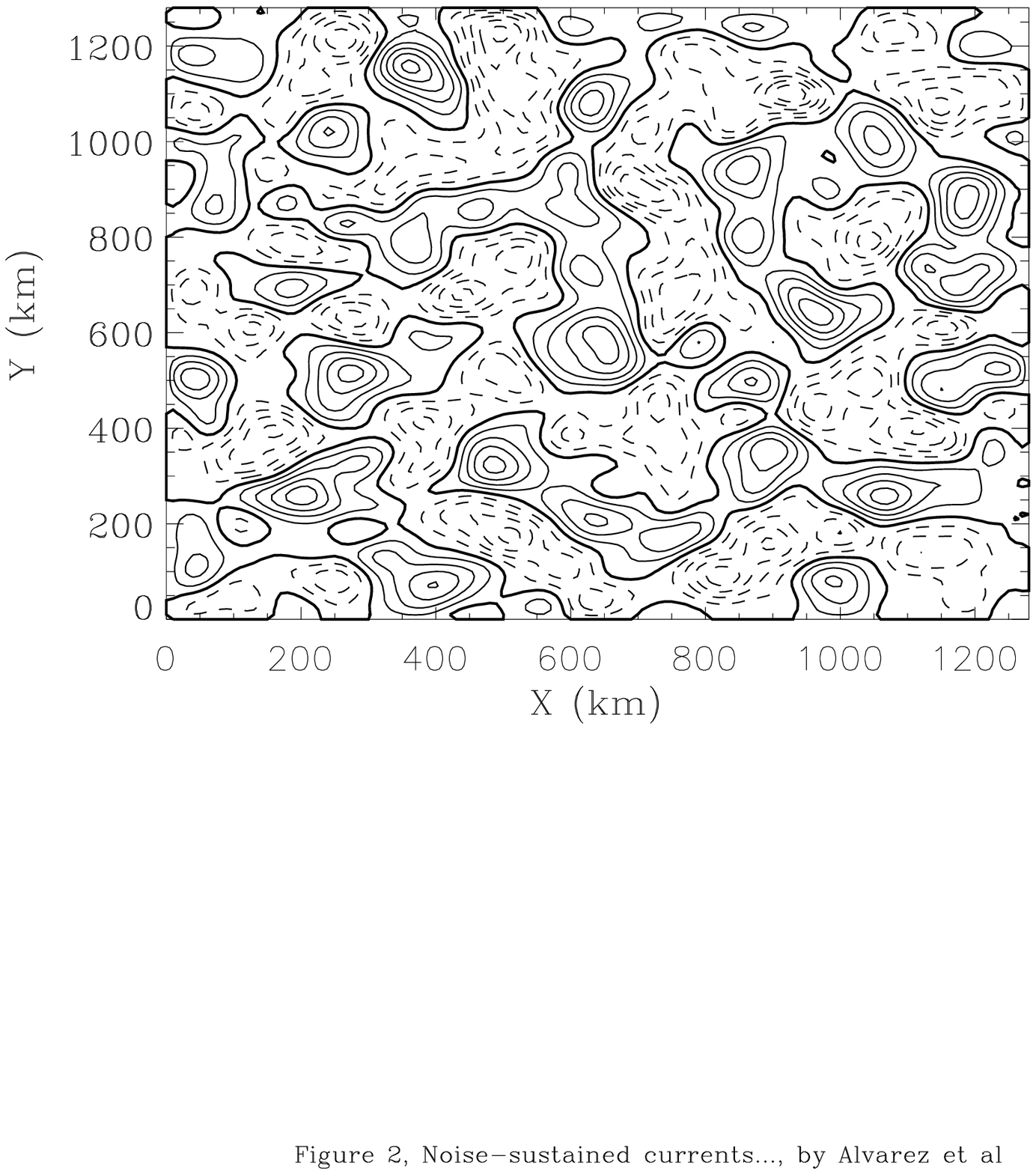}
\newpage
\epsfxsize=14cm
\epsfysize=14cm
\epsffile{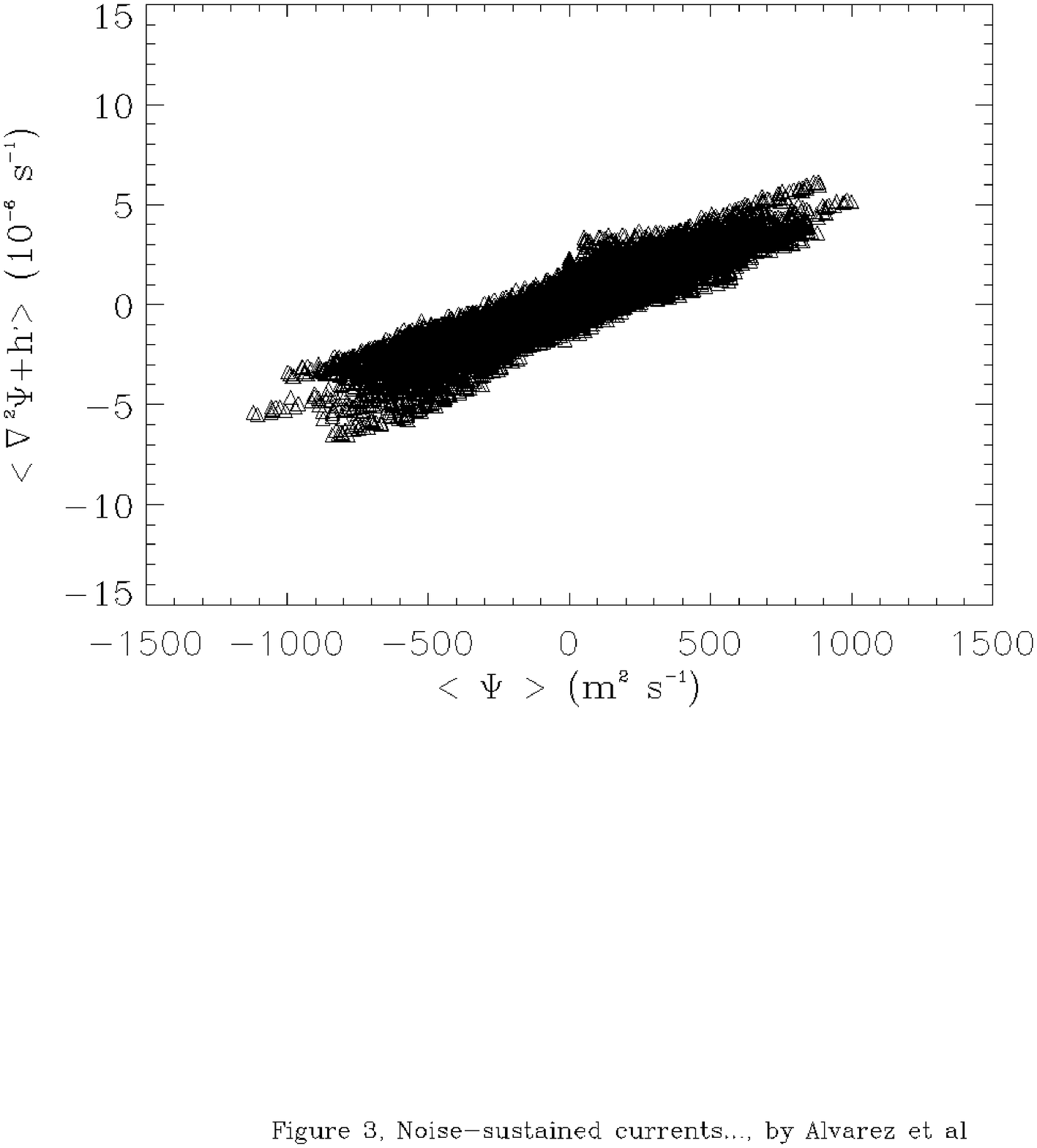}
\newpage
\epsfxsize=14cm
\epsfysize=14cm
\epsffile{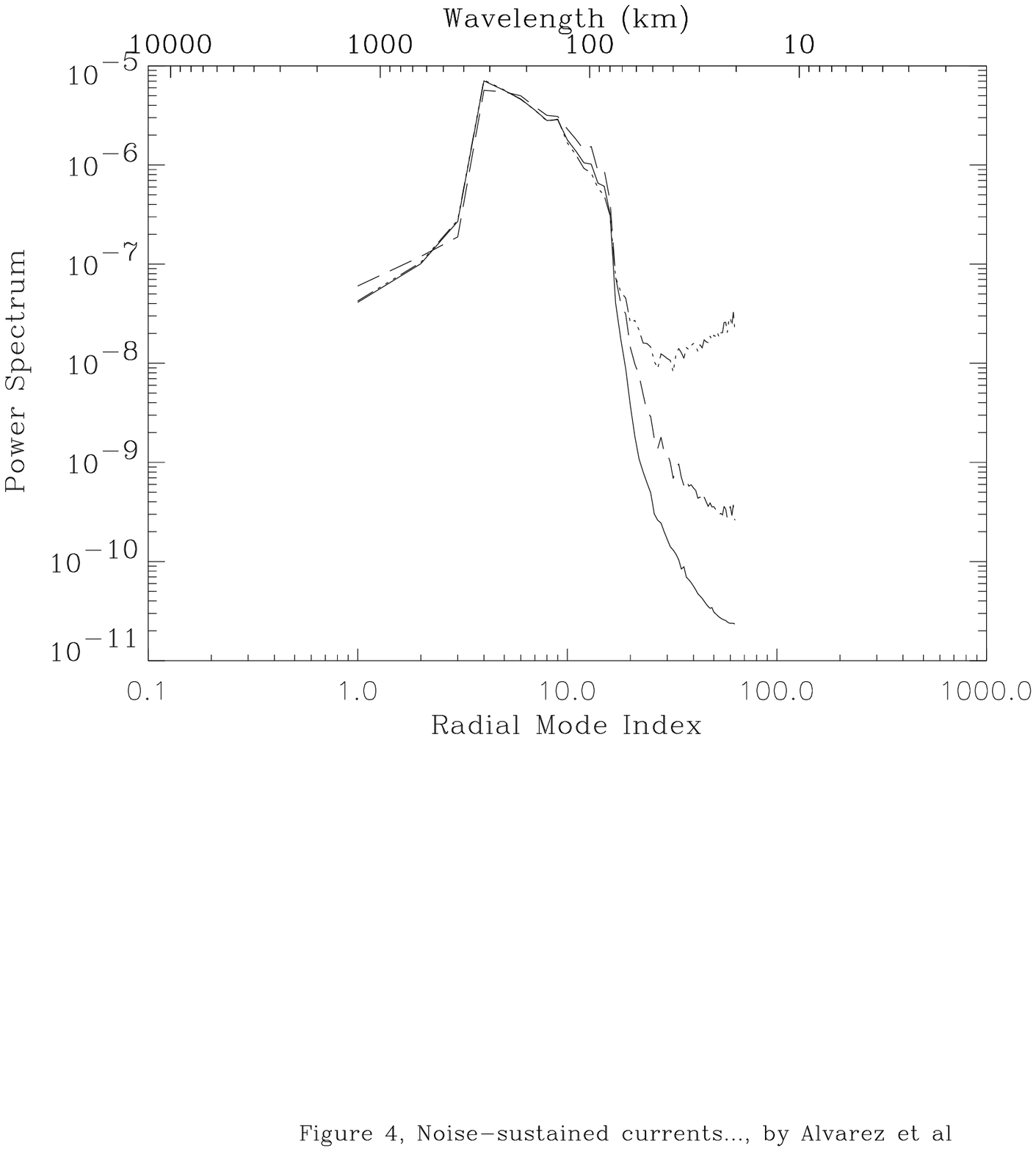}

\end{document}